\documentclass[conference]{IEEEtran}
\IEEEoverridecommandlockouts

\usepackage{cite}
\usepackage{amsmath,amssymb,amsfonts}
\usepackage{algorithmic}
\usepackage{graphicx}
\usepackage{textcomp}
\usepackage{xcolor}
\usepackage{textcomp}

\def\BibTeX{{\rm B\kern-.05em{\sc i\kern-.025em b}\kern-.08em
    T\kern-.1667em\lower.7ex\hbox{E}\kern-.125emX}}
\begin{document}

\title{MambaRate: Speech Quality Assessment Across Different Sampling Rates\\
  
}

\author{\IEEEauthorblockN{Panos Kakoulidis}
\IEEEauthorblockA{\textit{Innoetics} \\
\textit{Samsung Electronics}\\
Athens,Greece \\
p.kakoulidis@samsung.com}
\and
\IEEEauthorblockN{Iakovi Alexiou}
\IEEEauthorblockA{\textit{Innoetics} \\
	\textit{Samsung Electronics}\\
	Athens,Greece \\
	i.alexiou@partner.samsung.com}
\and
\IEEEauthorblockN{Junkwang Oh}
\IEEEauthorblockA{\textit{Mobile Communications Business} \\
	\textit{Samsung Electronics}\\
	Suwon, Republic of Korea \\
	jun1.oh@samsung.com}
\and
\IEEEauthorblockN{Gunu Jho}
\IEEEauthorblockA{\textit{Mobile Communications Business} \\
	\textit{Samsung Electronics}\\
	Suwon, Republic of Korea \\
	gunu.jho@samsung.com}
\and
\IEEEauthorblockN{Inchul Hwang}
\IEEEauthorblockA{\textit{Mobile Communications Business} \\
	\textit{Samsung Electronics}\\
	Suwon, Republic of Korea \\
	inc.hwang@samsung.com}
\and
\IEEEauthorblockN{Pirros Tsiakoulis}
\IEEEauthorblockA{\textit{Innoetics} \\
	\textit{Samsung Electronics}\\
	Athens,Greece \\
	p.tsiakoulis@samsung.com}
\and
\IEEEauthorblockN{Aimilios Chalamandaris}
\IEEEauthorblockA{\textit{Innoetics} \\
	\textit{Samsung Electronics}\\
	Athens,Greece \\
	aimilios.ch@samsung.com} 
}

\maketitle

\begin{abstract}
We propose MambaRate, which predicts Mean Opinion Scores (MOS) with limited
bias regarding the sampling rate of the waveform under evaluation. It is designed for Track 3 of
the AudioMOS Challenge 2025, which focuses on predicting MOS for speech in high sampling
frequencies. Our model leverages self-supervised embeddings and selective state space
modeling. The target ratings are encoded in a continuous representation via
Gaussian radial basis functions (RBF). The results of the challenge were based on the
system-level Spearman's Rank Correllation Coefficient (SRCC) metric. An initial MambaRate version 
(T16 system) outperformed the pre-trained baseline (B03) by $\sim$14\% in a few-shot setting without 
pre-training. T16 ranked fourth out of five in the challenge, differing by $\sim$6\% from the winning 
system. We present additional results on the BVCC dataset as well as ablations with different 
representations as input, which outperform the initial T16 version.
\end{abstract}


\section{Introduction}
Although subjective evaluation remains the gold standard for assessing speech-related models, 
relying only on human ratings is not scalable. As a solution, objective model-based methods, 
such as automatic MOS prediction systems, have been developed to predict human 
MOS ratings on given speech samples.  Most state-of-the-art non-intrusive MOS predictors focus on 
datasets of 16kHz \cite{MOSNet}, \cite{DNSMOS}, \cite{deepMOS}. This results in limited 
generalizability of these models when applied to higher-fidelity studio speech or mixed-bandwidth 
audio, affecting prediction accuracy. In real-world scenarios, speech data come at multiple 
sampling rates and varying audio quality.

Self-supervised-learning (SSL)-based approaches have been proven highly effective for MOS 
prediction tasks, showing superior performance and generalization compared to traditional 
features \cite{EricaCooper}, \cite{Comparison}. Recent work has leveraged SSL models such as 
wav2vec 2.0 \cite{wav2vec2}, HuBERT \cite{hubert}, and WavLM \cite{wavlm} to extract robust 
speech representations that, when fine-tuned, outperform  mel-frequency cepstrum coefficients (MFCCs) and spectrograms in modeling. Early models 
like MOSNet \cite{MOSNet} and MBNet \cite{mbnet} incorporated deep architectures and listener 
bias modeling, while fusion methods \cite{fusion}, combined multiple SSL encoders to further 
enhance prediction accuracy. More recently, lightweight architectures, such as SALF-MOS 
\cite{salfmos}, propose a speaker-agnostic approach, by downsampling latent features without 
the need to finetune or ensemble the SSL model, demonstrating strong generalization with 
minimal computational cost.

Recent benchmarks and challenges, such as the VoiceMOS Challenge \cite{voicemos}, have advanced non-intrusive 
MOS prediction under various acoustic distortions and language conditions. However, the 
challenge of predicting MOS for high-sampling frequencies has not been systematically 
studied. Addressing this limitation is important for deploying MOS predictors that achieve 
generalization in real-world scenarios and increasing resilience to sampling rate bias. 
Ultimately, developing robust MOS predictors will lead to the optimisation of objective 
evaluation methods, and thus reduce human intervention in large-scale speech evaluation.

\subsection{AudioMOS Challenge 2025 - Track 3}

The AudioMOS Challenge 2025 features three tracks. MambaRate's first version (T16) competes in Track 3, where a training set is 
provided consisting of 400 audio files with 16kHz, 24kHz or 48kHz sampling rate along with MOS ratings. 
These ratings derive from 10 listeners in 3 listening tests, each one including waveforms of the same sampling rate. 
There is no further information about the systems that generated these audios or what their target tasks were, 
e.g. voice conversion. The baseline \cite{baseline} is built on a finetuned wav2vec model with an additional 
linear layer, which processes the mean pooled SSL embeddings for MOS prediction. The systems are assessed with 
unseen audios and ratings from listening tests, which contained waveforms of all three sampling frequencies.

\section{Methodology}

\begin{figure}[t]
	\centering
	\includegraphics[width=30mm]{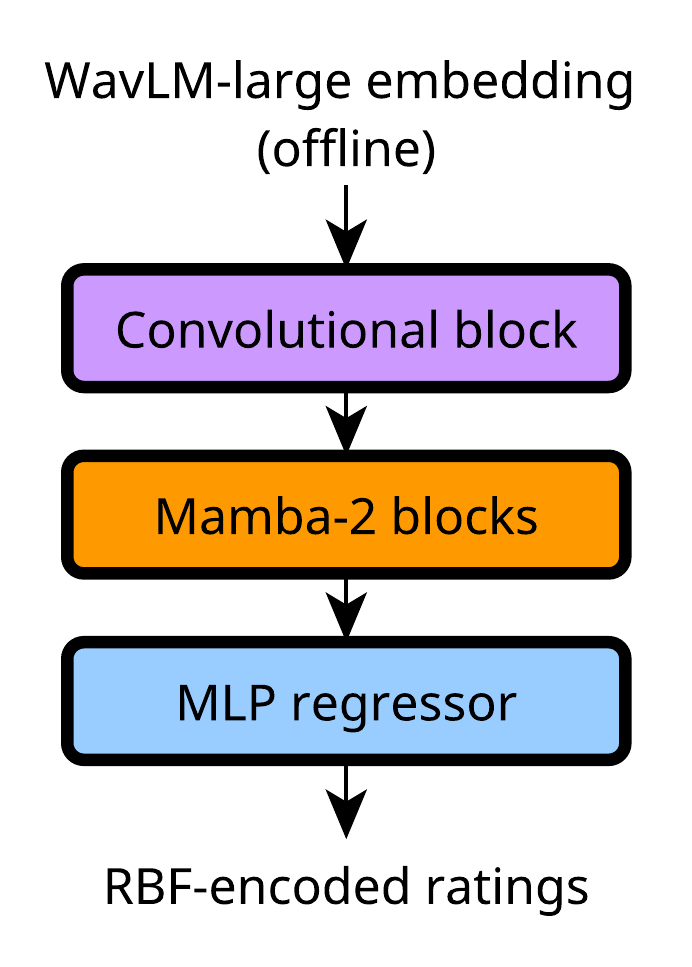}
	\caption{MambaRate's architecture}
	\label{fig:mambarate}
\end{figure}

We present MambaRate, which is a lightweight solution that generalizes across different sampling rates. 
There is no requirement for any type of metadata, such as listener or system identifiers. The model 
utilizes only SSL features, without the need for finetuning a large pre-trained SSL model. On the contrary, the baseline includes the parameters of the SSL model during its training. Although it strips 
some parts of the SSL model that are not needed for finetuning, it remains a large 
model overall, which requires hours of training. MambaRate is trained for a few minutes and for a few epochs 
($<$30), as imposed by an early stopping criterion. The overview of the model's architecture 
is illustrated in Figure \ref{fig:mambarate}.

MambaRate receives WavLM-Large \cite{wavlm} embeddings, which are extracted offline using 
publicly available weights \cite{huggingface}. For the T16 version, we retrieve representations from 
the sixth layer of the WavLM model, a layer that leads to better audio classification, compared to
the last layer of the stack \cite{voicerecood}. The embeddings are downsampled by a convolutional 
block consisting of an 1D convolutional layer (output=64 dim., kernel=3, stride=1), layer normalization 
and Mish activation \cite{mish}. A series of Mamba-based blocks, which 
are inspired by \cite{mamba4rec}, process the previous block's outputs. Each such block contains a Mamba-2 layer \cite{mamba2} ($d_{model}=64, 
d_{state}=32, d_{conv}=4, expand=8$) and a feedforward layer with Mish activation  \cite{mish}. The 
predictions are generated by a multi-perceptron block, having 2 linear layers. The last linear layer 
receives an output, which is averaged on the sequence length, and Sigmoid activation is applied on 
its outputs. In total, the model comprises almost 900k parameters, trained with AdamW optimizer 
\cite{adamw} (learning rate=0.001, weight decay=0.01), cosine annealing scheduler ($T_{max}=10$) and 
early stopping (patience=10, mindelta=0.001).

The target ratings are converted to continuous representations with a set of Gaussian 
radial basis functions defined as \cite{pocketminer}: 
\begin{equation}
	f(x, x_{c}) = exp\Big(- \frac{\| x - x_{c}  \|^2}{\sigma^2}\Big)
\end{equation}
where $x_{c}$ is a center. We use 16 centers to encode one rating value, therefore 
MambaRate's objective is to predict 16-dimensional vectors with the minimum mean squared error (MSE).
The values of the centers are evenly spaced in the ranking range (1-5). For example, the second 
center's value is $min + (max-min)/(centers - 1)$, where $min=1$,  $max=5$ and $centers=16$. Noise 
of 0.0001 is applied to the rating values before encoding to increase the robustness of the model. 
The predictions are decoded back to scalar values by selecting the center with the maximum value 
in the 16-dimensional vector for each utterance.

\section{Experiments and Results}

\subsection{Setup}

In our experiments we use the publicly available BVCC dataset \cite{bvcc} and the training data 
provided for Track 3 of the AudioMOS 2025 Challenge. All audios are preprocessed by utilizing the sv56 
pipeline for amplitude normalization as in \cite{baseline}, and downsampled to 16kHz for compatibility 
with the pre-trained models in use. The initial experiments use the median values of the ratings by utterance, 
since multiple sampling rates are involved. Our submitted T16 version is trained on these targets with 
a 90\%-10\% challange data split. The presented ablations are trained on mean ratings, as later 
experiments show that mean values greatly improve the models' performance. The metrics 
assigned by the challenge are mean squared error (MSE), linear correlation coefficient (LCC), Spearman's 
rank correlation coefficient (SRCC) and Kendall's tau (KTAU). We adopt these metrics for all experiments, 
whether at the utterance or system level.

\subsection{Comparisons with the baseline involving one sample rate}

We investigate how MambaRate performs in a conventional setting of MOS prediction. All the ratings 
come from BVCC listening tests that contain audios at the same sampling rate (16KHz). We use the 
publicly available weights of the baseline on BVCC for inference. We compute the utterance-level
metrics based on the open-source code of the baseline (Table \ref{tab1}). MambaRate slightly outperforms 
the baseline on BVCC dataset on MSE and KTAU metrics. The two compared models have almost identical LCC 
and SRCC values in this context. 

\begin{table}[htbp]
	\caption{Utterance-level evaluation for BVCC}
	\begin{center}
		\begin{tabular}{l|l|l|}
			\cline{2-3}
			& Baseline                          & MambaRate                          \\\hline 
			\multicolumn{1}{|l|}{MSE \textdownarrow{}} & \multicolumn{1}{l|}{0.277}        &       \multicolumn{1}{l|}{ \textbf{0.209}}                \\ \hline
			\multicolumn{1}{|l|}{LCC \textuparrow{}} & \multicolumn{1}{l|}{\textbf{0.869}}           &      \multicolumn{1}{l|}{0.865}  \\ \hline
			\multicolumn{1}{|l|}{SRCC  \textuparrow{}} & \multicolumn{1}{l|}{\textbf{0.869}}          &        \multicolumn{1}{l|}{0.866}     \\ \hline
			\multicolumn{1}{|l|}{KTAU  \textuparrow{}} & \multicolumn{1}{l|}{0.690}   &      \textbf{0.736}      \\ \hline
		\end{tabular}
		\label{tab1}
	\end{center}
\end{table}

\subsection{Challenge-related experiments with three sampling rates}

\begin{figure}[t]
	\centering
	\includegraphics[width=70mm]{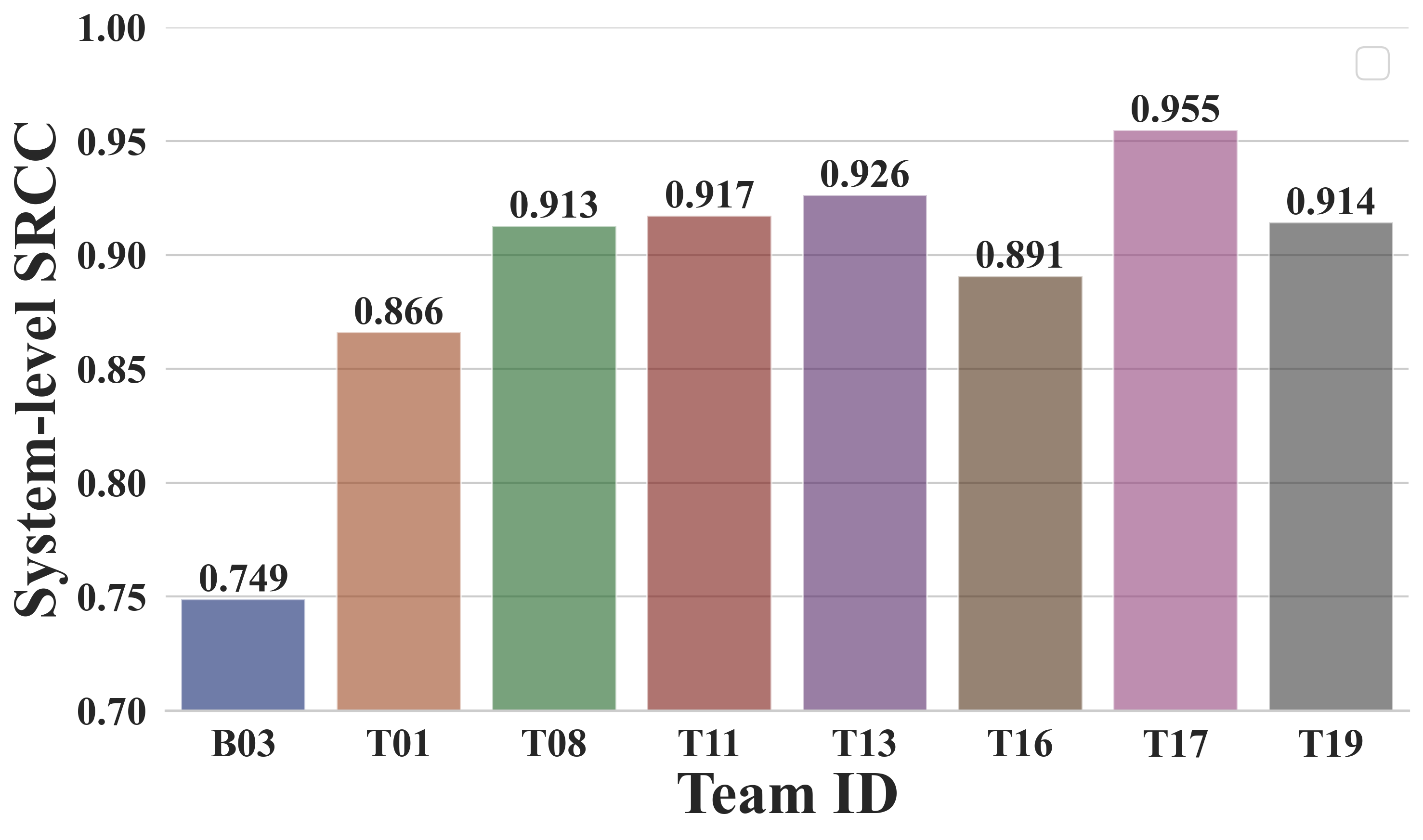}
	\caption{The initial version of MambaRate (T16) in the AudioMOS 2025 results for Track 3}
	\label{fig:track3}
\end{figure}

Firstly, we compare the baseline with MambaRate on the challenge's data (Table \ref{tab2}). 
The zero-shot inference of the baseline underperforms, while the fine-tuned baseline demonstrates 
great improvement. However, MambaRate clearly outperforms both setups of the baseline, although 
the former is trained only on the training data of the challenge. T16 version is trained on a different 
split of the training data (90\% for training), gaining the fourth place in the challenge 
(Figure \ref{fig:track3}). Also, we notice that pre-training our model on BVCC leads to refined predictions.

\begin{table}[htbp]
	\caption{Utterance-level evaluation for experiments}
	\begin{center}
		\begin{tabular}{l|ll|ll|p{6cm}}
			\cline{2-5}
			& \multicolumn{2}{c|}{Baseline}                            & \multicolumn{2}{c|}{MambaRate}                            \\ \cline{2-5} 
			& \multicolumn{1}{c|}{BVCC} & \multicolumn{1}{c|}{BVCC+C} & \multicolumn{1}{c|}{C} & BVCC+C \\ \hline
			\multicolumn{1}{|l|}{MSE  \textdownarrow{}} & \multicolumn{1}{l|}{1.069}   &        0.336                 &  \multicolumn{1}{l|}{0.086}   &            \textbf{0.071}             \\ \hline
			\multicolumn{1}{|l|}{LCC \textuparrow{}} & \multicolumn{1}{l|}{0.748}   &               0.854          & \multicolumn{1}{l|}{0.871}   &                \textbf{0.891}         \\ \hline
			\multicolumn{1}{|l|}{SRCC  \textuparrow{}} & \multicolumn{1}{l|}{0.522}   &              0.853           & \multicolumn{1}{l|}{0.879}   &         \textbf{0.930}               \\ \hline
			\multicolumn{1}{|l|}{KTAU  \textuparrow{}} & \multicolumn{1}{l|}{0.350}   &            0.686             & \multicolumn{1}{l|}{0.738}   &          \textbf{0.822}               \\ \hline
			\multicolumn{5}{p{6cm}}{$^{\mathrm{a}}$ BVCC: trained on BVCC, C: trained on challenge's training data, BVCC+C: pre-trained on BVCC and fine-tuned on challenge's training data. Metrics are computed by averaging ratings for all sampling rates per
				utterance. C and BVCC+C models are trained  on median ratings.}
		\end{tabular}
		\label{tab2}
	\end{center}
\end{table}

\subsection{Ablation studies on the challenge's data}

We experimented further with MambaRate using the challenge's training data (70-15-15 split). We share the 
insights of training the model with different pre-extracted embeddings as input for the learning of mean ratings. 
We explore more of the top performing WavLM layers presented in \cite{voicerecood} (Table \ref{tab3}). Our findings 
indicate that each layer has a complementary contribution, addressing different aspects of 
the prediction. Notably, embeddings from the 9th layer outperform a pre-trained model with the 
6th layer embeddings. We also train ablations that use raw embeddings from Whisper (large-v3-turbo)
\cite{whisper}, a state-of-the-art Automatic Speech Recognition (ASR) model, and from 
VGGish\cite{vggish}, an established audio classification model. Both types of representations 
demonstrate their own improvements over WavLM-Large (Table \ref{tab4}). They derive from the final 
outputs of each compared model and are extracted from publicly available weights \cite{huggingface, 
tensorflowmodelgarden}.

\begin{table}[htbp]
	\caption{Ablations with representations from different layers of WavLM-Large}
	\label{table:layers}
	\begin{center}
		\begin{tabular}{|c|c|c|c|c|c|p{8cm}}
			\hline
			Systems & & MSE  \textdownarrow{}& LCC  \textuparrow{}& SRCC  \textuparrow{}& KTAU  \textuparrow{} \\
			\hline
			WavLM6 & U & 0.057 & 0.922 & 0.899 & 0.782 \\
			\cline{2-6}
			&       S &   0.040   &  0.963   &  \textbf{0.976}   & \textbf{0.929} \\
			\hline
			WavLM6-P & U & 0.046 & 0.927 & 0.911 & 0.799 \\
			\cline{2-6}
			& S & 0.067 & 0.939 & 0.952  &  0.857 \\
			\hline
			WavLM7 & U & 0.057 & 0.906 & 0.869 & 0.732 \\
			\cline{2-6}
			&       S &   \textbf{0.017}   &  \textbf{0.982}   &  \textbf{0.976}   & 0.928 \\
			\hline
			WavLM8 & U & 0.054 & 0.911 & 0.893 & 0.766 \\
			\cline{2-6}
			&       S &  0.037   &  0.964   &  \textbf{0.976}   & 0.928 \\
			\hline
			WavLM9 & U & \textbf{0.038} & \textbf{0.937} & 0.911 & 0.801 \\
			\cline{2-6}
			&       S &   0.033   &  0.967   &  \textbf{0.976}   & 0.928 \\
			\hline
			WavLM10 & U & 0.057 & 0.908 & \textbf{0.919} & \textbf{0.812} \\
			\cline{2-6}
			&       S &   0.042   &  0.963   &  0.952   & 0.857 \\
			\hline
			WavLM & U & 0.048 & 0.924 & 0.897 & 0.770 \\
			\cline{2-6}
			& S &   0.062   &     0.938   &    0.905  &  0.786  \\

			\hline
			\multicolumn{6}{p{8cm}}{$^{\mathrm{a}}$ U: utterance-level, S: system-level, 
				"WavLM[N]" : embeddings extracted from Nth layer, "-P": pre-trained on BVCC.
				 Metrics are computed by averaging ratings for all sampling rates per utterance/system. 
				 Models are trained  on mean ratings}
		\end{tabular}
		\label{tab3}
	\end{center}
\end{table}

\begin{table}[htbp]
	\caption{Ablations with representations from different models}
	\label{table:models}
	\begin{center}
		\begin{tabular}{|c|c|c|c|c|c|p{8cm}}
			\hline
			Systems & & MSE  \textdownarrow{}& LCC  \textuparrow{}& SRCC  \textuparrow{}& KTAU  \textuparrow{} \\
			\hline
			WavLM & U & 0.048 & 0.924 & 0.897 & 0.770 \\
			\cline{2-6}
			& S &   0.062   &     0.938   &    0.905  &  0.786  \\
			\hline
			Whisper & U &  \textbf{0.042} &  \textbf{0.929} & \textbf{0.925} & \textbf{0.816} \\
			\cline{2-6}
			& S &    \textbf{0.037}  &   \textbf{0.963}  &   0.905  &  0.786  \\
			\hline
			VGGish & U & 0.082 & 0.856 & 0.813 & 0.686 \\
			\cline{2-6}
			& S &    0.043  &   0.955   &  \textbf{0.952}  & \textbf{0.857}  \\
			\hline
			\multicolumn{6}{p{8cm}}{$^{\mathrm{a}}$ U: utterance-level, S: system-level.
				Metrics are computed by averaging ratings for all sampling rates per utterance/system. 
				Models are trained  on mean ratings.}
		\end{tabular}
		\label{tab4}
	\end{center}
\end{table}

\section{Conclusion and future work}

We present MambaRate, a model that outperforms the baseline of Track 3 in the
AudioMOS Challenge 2025. The submitted T16 ranks fourth out of five, lying close to the winning system. We showcased that offline pre-trained representations can be efficiently adopted in a fewshot manner for MOS prediction, supporting different sampling rates . In the ablation studies, we explore the usage of embeddings from different layers of the chosen SSL model. We also consider representations from other large models 
for learning mean ratings. These ablations outperform T16 and lay the foundation for future work. Efficiently combining the best performing layers of the large pre-trained models may lead to more 
accurate predictions. Testing MambaRate on other public datasets, such as SOMOS \cite{somos} and 
VCC2018 \cite{vcc2018}, is another essential direction we plan to pursue. Enriching such datasets 
with upsampled versions of their included waveforms, e.g. via neural bandwidth extension, and conducting 
additional listening tests with these augmented versions, may offer a promising solution for this task.

\bibliographystyle{IEEEtran}
\bibliography{mybib}

\end{document}